\documentclass[article,nofootinbib,amsmath,superscriptaddress,tightenlines,floats,floatfix]{revtex4}
\usepackage{graphicx}
\usepackage{bm}
\usepackage{epsfig}
\usepackage{color}
\usepackage{colordvi}
\begin{document}

\title{Scalar gravity: Post-Newtonian corrections via an effective field theory approach}

\author{Rafael A. Porto}

\affiliation{Department of Physics, Carnegie Mellon University, Pittsburgh, PA 15213
\vspace{0.3cm}}

\author{Riccardo Sturani}

\affiliation{Physics Department, Univ. Genva, 24 Quai E. Ansermet, 1211 Geneve,
Switzerland and INFN, Presidenza dell'INFN, Roma, Italy \vspace{0.3cm}}

\begin{abstract}
The problem of motion in General Relativity has lost its academic status and
become an active research area since
the next generation of gravity wave detectors will rely upon its solution. Here we
will show within scalar gravity, how ideas borrowed from Quantum Field Theory can be used to solve the problem
of motion in a systematic fashion.
We will concentrate in Post-Newtonian corrections.
We will calculate the Einstein-Infeld-Hoffmann action and show how a systematic perturbative expansion
puts strong constraints on the couplings of non-derivative interactions in the theory.
\end{abstract}

\maketitle

\section{Introduction}

Consider the apparently simple problem of the earth motion around
the sun. The Newtonian solution is an excellent approximation but suppose we wish to
be more accurate. A closer look reveals that there are many
sources of complication. Einstein theory teaches us how to correct
for relativistic effects. However, the earth is clearly not a
point particle, and will thus deform under the influence of tidal
forces. In addition the whole sun-earth system will radiate energy
away in the form of gravitational waves. The
inclusion of all these effects can make the problem of solving for
the trajectory intractable. In the past, solving this problem was
only of academic interest, but the next generation of gravity wave
detectors will rely upon its solution \cite{ligo}. The
construction of accurate templates for gravity wave
interferometers is a daunting task. After more than ten years of
work the templates have been completed up to third Post-Newtonian
(PN) order for non-spinning compact binaries \cite{damour}.
However, it was not clear how to: proceed to higher
orders in a systematic fashion, include finite size effects
due to spin or spin-spin corrections. During the last years a new
framework has emerged, coined NRGR (Non-Relativistic General
Relativity) due to its similarities with Effective Field Theory
(EFT) ideas in particle physics, where all of the apparent
obstacles of the traditional approach can been successfully
overcome \cite{nrgr,spin}. NRGR naturally allows for a systematic
account of the internal structure of the binary constitutes and
permits us to calculate back reaction as well as dissipative effects
\cite{nrgr3}. Moreover, new results for spinning compact binaries
have been recently reported \cite{eih}. In this short contribution
we will show within scalar gravity, how an EFT
approach can be used to solve the problem of motion in a
systematic fashion \cite{walter}. In particular we will calculate
the Einstein-Infeld-Hoffmann (EIH) action for the case of two {\it
scalar-gravitating} bodies, accurate up to 1PN. The purpose of
this contribution is pedagogical, allowing us to concentrate on
the conceptual aspects. As we shall see a systematic perturbative expansion
puts strong constraints on the couplings of non-derivative interactions in the theory.

\section{Scalar gravity}

The starting point of the EFT approach consists of a theory of point particles
coupled to a real scalar field $\phi$ we shall call the ``s-graviton''.
For simplicity we will consider here a massive $\phi^3$ theory in a Minkowski background,
though we will discuss other type of models later on. The action will be given by $S = S_g + S_{pp}$, with
\begin{equation}\label{sg}
S_g = \int d^4x ~\left(\partial_\mu \phi \partial^\mu \phi -\mu^2\phi^2 - \lambda \phi^3\right), \;\;\;
S_{pp} = -\sum_a m_a \int d\tau_a\sqrt{1+\frac{\phi}{M}}
\end{equation}
describing the s-graviton dynamics and motion of the binary system
($a=1,2$). In this equation $M$ sets the coupling to matter, and
$\lambda,\mu$, the self-interaction and s-graviton mass
respectively. Also $d\tau =\sqrt{\eta^{\mu\nu}dx^\nu dx^\mu}$
represents the proper time along the $a$-th particle and
$\eta^{\mu\nu} \equiv diag(+,-,-,-)$, we work in $\hbar=c=1$
units. The choice of matter coupling is meant to resemble Einstein
case, at least for the $h_{00}$ mode, with $M$ playing the role of
the Planck Mass. The normalization is also chosen to mimic the
graviton propagator. We could in principle add a set of higher
order operators in the worldline action to account for finite size
effects. However, $\phi^3$ theory is super-renormalizable and it
is possible to show that the $n$-point function is UV finite and
no higher order operators are needed \footnote{Notice also that
using field re-definitions ($\sim$ equations of motion) we can
always trade $\partial^2 \phi$ by a polynomial, so higher
dimensional operator are always of the form $\phi^\alpha$ and could be absorbed
into the worldline couplings.}.
One other aspect of the super-renormalizability is the fact that a
$\phi^3$ self-interaction in four spacetime dimensions has a
dimensionful coupling and the perturbative approach breaks
down at distances of order $1/\lambda$. This is connected to IR
divergences (in the massless limit) which appear in the
perturbative expansion due to factors of $\lambda/E$, with $E$ the
energy of the s-graviton. These IR divergences must cancel in any
physical observable, such as the binding energy of the binary
system. However, a resummation procedure is in general needed in
order to achieve a finite result \cite{jackiw}. There are a few
ways to overcome this. We could work in six dimensions where $\lambda$ is dimensionless,
or with a IR cutoff. Instead we adopted a small s-graviton mass. Notice that a
mass term can be produced by a tadpole mechanism, therefore a
s-graviton mass, $\mu$, would be $\it naturally$ generated by
quantum fluctuations since no symmetry prevents it. One would then
expect $\mu \sim \lambda$. We will see in what follows how a
well defined perturbation theory puts strong
constraints in the self-interaction coupling of the theory. We
will discuss later on under which circumstances this is a more
generic phenomena.

\section{NRGR}

The power of the EFT formalism resides in a manifest power counting in the expansion parameter of the theory,
in this case the relative velocity $v$. Here we will pinpoint the necessary steps and refer to Goldberger's contribution
for further details \cite{walter}. The expansion of the worldline Lagrangian leads to
\begin{equation}
L_{pp} = \sum_a \frac{m_a}{2} \left[{\bf v}_a^2 -\left(1 - \frac{{\bf v}_a^2}{2}\right)\frac{\phi}{M}+\frac{1}{4}{\bf v}_a^4 + \frac{1}{4}\frac{\phi^2}{M^2}\right]+... ,
\end{equation}
where we have chosen $x^0$ as the worldline parameter.
The propagator for the field $\phi$ appearing in $L_{pp}$ is still fully relativistic,
and therefore a small velocity expansion has yet to be performed. To deal with
this problem it is convenient to decompose the s-graviton field into potential modes ($\bar\phi$) with momentum scaling
$k^\mu \sim (v/r,1/r)$  (notice they can never go on shell), and radiation modes ($\Phi$)
whose momentum scale as $k^\mu \sim (v/r,v/r)$. In the EFT spirit potential modes
do not propagate and can be thus integrated out at each order in
perturbation theory. Radiation s-gravitons on the other hand can appear
on shell and must be kept as propagating degrees of freedom in
order to reproduce the correct long distance physics.

\subsection{Power counting}

In the EFT approach one computes the effective action perturbatively, in
our case in $v$, based on systematic power counting rules. In
order to obtain the latter one starts with the scaling laws for
the $(\bar \phi,\Phi)$ fields. For convenience one first introduces $\Phi_{\bf k}$, where the large
momentum piece of the potential s-graviton is factored out \cite{nrgr,walter}. By expanding (\ref{sg}) we get
\begin{equation}
\langle\Phi_{\bf k}(x^0)\Phi_{\bf q}(0)\rangle =  (2\pi)^3\delta^3({\bf k}+{\bf q})\delta(x^0)\frac{-i}{2({\bf k}^2+\mu^2 )}\label{p1}, \;\;\;\;
\langle\bar\phi(x)\bar\phi(0)\rangle = \int \frac{d^4k}{(2\pi)^4} \frac{i}{2(k^2+\mu^2)} e^{ikx}
\end{equation}
for the propagators. Notice that we have decided to keep the mass
``non-perturbatively'' to cure IR divergences, although we will
assume $\mu r < v$ in what follows, and treat it as a perturbation
when allowed, in order to resemble the massless power counting
rules and a $1/r$ leading order potential. If we assign the
scaling $x^0 \sim v/r$ we obtain the following leading order power counting rules
\begin{equation}
\bar\phi \sim v/r, \;\;\; \Phi_{\bf k} \sim \sqrt{v}r^2 \to \Phi \sim M\frac{v^2}{\sqrt{L}},
\end{equation}
where $L = mvr$. The last arrow follows from the assumption
that the leading order potential is given by $1/r$ and hence the
virial theorem, $v^2 \sim \frac{m}{M^2r}$, applies
\footnote{Notice it also implies $m/M \sim \sqrt{Lv}$}. This
assumption is true in the case of $\lambda=0,\mu r < v$,
however $\lambda \neq 0$ can easily jeopardize our power
counting due to the introduction of a dimensionful coupling. In
order to tackle this problem let us exercise our scaling rules and
power count the first correction due to $\lambda$. The diagram is shown in fig. 2a and it scales as
\begin{equation}
\mbox{fig.~2a} \sim \left[dx^0\frac{m}{M}\Phi\right]^3\left[dx^0 \delta^3({\bf k})\lambda \Phi^3\right] \sim
\left[\frac{r}{v}\frac{m}{M}M\frac{v^2}{\sqrt{L}}\right]^3\frac{r}{v}\frac{\lambda}{r^3}\left[M\frac{v^2}{\sqrt{L}}\right]^3 \sim L v^2 (\lambda M r^2),
\end{equation}
$v^2(\lambda M r^2)$ times the leading order term which scales as $L$ \cite{nrgr}.
It is easy to see higher order terms in $\lambda$ follow the same pattern. For $\lambda=g^2M$, with $g$ a dimensionless
coupling, we end up with $r < \frac{1}{gM}$ for the validity of the perturbative approximation
and power counting. In order to make sense of the perturbative approach we had to cure this IR
singularities before expecting any power counting to work, and
that is what the s-graviton mass is doing. If we demand our leading
order potential to match the Newtonian case we will set $M \equiv
m_{Pl}$ and therefore the perturbative expansion is valid for $r < l_{Pl}/g$, with $l_{Pl}=10^{-33}$ cm, the Planck length.
To avoid entering the quantum realm\footnote{Recall that loop effects in
NRGR for gravitons are suppressed by $1/L$ and can be thus ignored in the classical
scenario \cite{nrgr}.} we will have to fine tune $g$ to an
extremely small number of the order of $10^{-40}$ for typical
binary systems in the solar mass range in the inspiral regime.
This obviously defies naturalness arguments and puts a flag on the
phenomenological viability of such theory since it implies a ridiculously small self-coupling, $\lambda \sim 10^{-80} m_{Pl}\sim
10^{-60} GeV$!
Notice that the problem does not lie in the self-coupling itself but in the strength
of the worldline coupling which determines the leading order scaling laws.
In Einstein theory this is taken care of by the three graviton coupling, $g_3 \sim {\bf k}^2/M \to g_3 M
r^2 \sim 1$. The condition $\mu \leq v/r$ also implies a
stunningly tiny s-graviton mass of the order of $10^{-30} ev$. These
are consistent, and somehow equivalent, to solar system
constraints \cite{mass}, whereas by naturalness arguments $\mu
\sim \lambda$ would produce an even smaller value. We will hereon
assume $\mu r \sim v, \lambda r \ll 1$, and proceed with this
theory as a playground.

\section{Einstein-Infeld-Hoffmann}

Let us concentrate now in the calculation of the 1PN correction
to the gravitational potential. The leading order one s-graviton
exchange can be easily seen to reproduce Newtonian gravity
\cite{nrgr}. We also need to take into account diagrams with one single
s-graviton exchange which are down by $v^2$ shown in figures 1a
and 1b plus the non linear terms depicted in figures 2a and 2b. We
will treat the s-graviton mass as perturbation in the one s-graviton
exchange ($\mu r \sim v$) and that is shown in diagram 1c. The
computation proceeds systematically by using the Feynman rules of
the EFT order by order. We will concentrate in detail in fig. 2a, we
will display the full result later on. For the three
s-graviton diagram we will have
\begin{equation}
\mbox{fig.~2a} = \frac{1}{2}\left(\frac{-im_2}{2M}\right)^2\frac{-im_1}{2M}\int dt_1 dt_2 dt_{2'} \langle T(\Phi (x_1)\Phi (x_2)\Phi (x_{2'}))\rangle\label{3g}.
\end{equation}

Our task now is to compute the three-point function. For a $\phi^3$ theory one obtains
\begin{equation}
\langle T(\Phi (x_1)\Phi (x_2)\Phi (x_{2'}))\rangle = 3! (-i\lambda) \delta(t_1-t_2)\delta(t_1-t_{2'})
\int \prod_r^3 \frac{d^3{\bf k}_r}{(2\pi)^3} e^{-i\sum_i {{\bf k}_i}\cdot{{\bf x}_i}}
(2\pi)^3\delta^3\left(\sum_i^3 {{\bf k}_i}\right) \prod_j^3 \frac{-i}{2({\bf k}_j^2+\mu^2)}\label{t}.
\end{equation}

The next step would be to plug this expression back into
(\ref{3g}), get a finite result which we will have to further
expand in powers of $\mu r \sim v$ and keep the leading order
piece, already at 1PN for $ \lambda M r^2 \sim 1$. In the EFT
spirit a better way to proceed is to treat $\mu$ as a
perturbation in the same way time derivatives are treated, by
expanding the propagators in powers of $\mu/|{\bf k}|$. For the
one s-graviton exchange this represents no harm. In general one faces the
problem that IR divergences will only cancel out after all the terms are included.
If we are willing to accept that is the case one can calculate the 1PN correction
by taking the massless limit of (\ref{t}) and keep the (non-constant) finite
piece. Therefore, introducing $d=3+\epsilon$ and taking the limit $\epsilon \to 0$ one gets
\begin{eqnarray}
\mbox{fig.~2a} &=&  i \lambda \frac{3m_1m_2^2}{64M^3} \int dt\frac{d^3{\bf k}_2}{(2\pi)^3}\frac{d^3{\bf k}_1}{(2\pi)^3} \frac{1}{{\bf k}_1^2{\bf k}_2^2({\bf k}_1+{\bf k}_2)^2}
e^{-i{\bf k}_1\cdot ({\bf x}_1-{\bf x}_2)} = i \lambda \frac{3\pi G_N m_1m_2^2}{16M} \int dt \frac{d^d{\bf k}_1}{(2\pi)^3} \frac{1}{({\bf k}_1^2)^{3/2}}e^{-i{\bf k}_1\cdot ({\bf x}_1-{\bf x}_2)}\nonumber\\
&=& i 3 \lambda \frac{G_N m_1m_2^2}{64\pi M} \Gamma(\epsilon/2)\int dt\left(\frac{{|\bf x}_1-{\bf x}_2|^2}{4}\right)^{-\epsilon/2} \to - i 3 \lambda M G_N^2 m_2^2m_1~\int dt\log (\mu |{\bf x}_1-{\bf x}_2|) +\mbox{constant}.
\end{eqnarray}
with $G_N=\frac{1}{32\pi M^2}$, and the ``constant" piece also
contains the $\frac{1}{\epsilon}$ IR pole\footnote{In the massive
case this is translated into a $\log\mu$ factor. The full result
is a Bessel function, $K_0(\mu r)$, whose leading order piece in
$\mu r$ reproduces the logarithmic potential.}.

\begin{figure}[!t]
\centerline{\scalebox{0.6}{\includegraphics{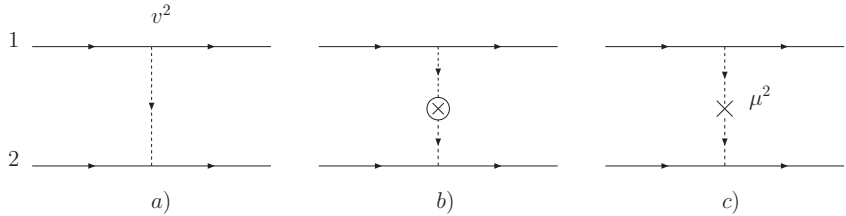}}}
\vskip-0.3cm \caption[1]{One s-graviton exchange contribution at
1PN. The $\bigotimes$ represents a correction to the propagator,
and $\times$ a mass insertion.}\label{1}
\end{figure}

\begin{figure}[!t]
\centerline{\scalebox{1.3}{\includegraphics{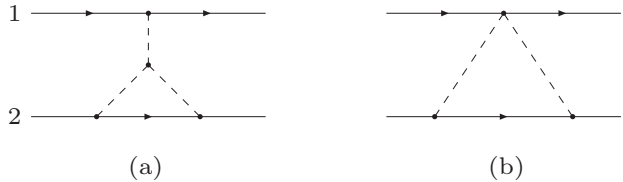}}}
\vskip-0.3cm \caption[1]{Non linear contributions at
1PN.}\label{2}
\end{figure}

Our final task consists in collecting the other few pieces. We refer to \cite{nrgr,walter} for details since the
calculations are almost identical. Let us compute the result for the new term in fig.~1c due to the s-graviton mass insertion,
\begin{equation}
\mbox{fig.~1c} = -i{m_1 m_2\over 8 M^2}\int dt_1 dt_2 \delta(t_1-t_2)
\int \frac{d^3{\bf k}}{(2\pi)^3} {\mu^2\over {\bf k}^4} e^{-i{\bf k}\cdot({\bf x}_1 -{\bf x}_2)}
={i\over 2}\int dt ~G_N m_1 m_2\mu^2 |{\bf x}_1 - {\bf x}_2|,
\end{equation}
which is nothing but the ${\cal O} (v^2)$ piece in the expansion
of the Yukawa potential, $-\frac{e^{-\mu r}}{r}
\sim \mu - \frac{1}{r} (1 + \frac{\mu^2 r^2}{2}+...)$.

Putting everything together, including mirror images, we finally obtain
\begin{eqnarray}
L_{EIH} &=&
\label{eih}
{1\over 8}\sum_a m_a {\bf v}^4_a + {G_N m_1 m_2\over 2 |{\bf x}_1 -{\bf x}_2|}\left[{\bf v}^2_1+ {\bf v}^2_2 +({\bf v}_1\cdot {\bf v}_2)  -
{({\bf v}_1\cdot  {\bf x}_{12} )({\bf v}_2\cdot {\bf x}_{12})\over |{\bf x}_1-{\bf x}_2|^2} \right]
+ {G^2_N m_1 m_2 (m_1+m_2)\over 2  |{\bf x}_1 -{\bf x}_2|^2}\nonumber\\
&+& \frac{1}{2}~G_N ~m_1 m_2 \mu^2 |{\bf x}_1 - {\bf x}_2| - 3\lambda ~G_N^2 ~M~m_2m_1(m_1+m_2)~\log (\mu |{\bf x}_1-{\bf x}_2|)
\end{eqnarray}
where we have also included the relativistic corrections to the
kinetic energy of the point particles. The logarithmic potential
introduces a very interesting feature, namely a $\frac{1}{r}$
force into the equations of motion and therefore $v^2 \sim a/r + b
+ ...$, which implies a {\it dark matter} type of effect for the
galaxy rotation curves. This is however by no means a serious
candidate and we mention this only as a curiosity.

\section{Discussion - Conclusions}

The EFT approach is a powerful tool within the PN framework. Using
no more than dimensional analysis many conclusions can be already
drawn before dwelling into the details of the calculations. We
applied the techniques in the case of a massive $\phi^3$ theory as
a playground but the ideas can be easily extended to more
complicated scenarios. From the NRGR power counting rules we learned that
in order to produce a well defined perturbative
expansion, $\lambda$ had to be fine tuned to a ridiculously small
(compared with $m_{Pl}$) scale. One might however ask whether this
is a feature of a $\phi^3$ theory or will this be faced in
other scenarios. Let us consider a more general case,
\begin{equation}
\label{ste}
S_{\phi} = \int d^4x ~\left(\gamma(\phi) \partial_\mu \phi \partial^\mu \phi + B(\phi)\right)
\end{equation}

In what follows we will consider two distinct case.

\subsection{$B(\phi)=0$}

 If we expand $\gamma(\phi) \sim 1 + \phi/M + ...$,  we will get a
kinetic piece plus a potential $V(\phi)$ with terms like
$\left(\frac{\phi}{M}\right)^n \phi\partial^2 \phi$ $n\geq 1$.
This theory is not renormalizable, and it is easy to show it
resembles Einstein case. We can also show that the perturbative
approach is under control by power counting the contribution from
a generic term in $V(\phi)$. For an $(n+2)$ s-graviton diagram we
will get,
\begin{equation}
\sqrt{L}^{n+2}\left(\frac{v^2}{\sqrt{L}}\right)^{n+2}M^2r^2/v \sim  L v^{2n}.
\end{equation}

For instance the first term in the expansion, $\phi^2 \partial^2
\phi$, resembles the three graviton coupling in Einstein theory
(up to tensor structure). Had we chosen this interaction we would
have ended up with a similar 1PN correction as in the original
Einstein-Infeld-Hoffmann action \cite{nrgr,walter}.

\subsection{$B(\phi) \neq 0$}

This case is substantially different. Let us study a generic term, $g M^4(\phi/M)^n$,
with $g$ a dimensionless coupling and $n \geq 4$. The
$n$ s-graviton diagram will scale as
\begin{equation}
g M^4r^4 v^{2n-1} \sim  g L^2 v^{2n-7} \sim L ~ g(m/M)^2 v^{2n-8}.
\end{equation}
To have a controlled perturbative expansion we would have to impose
$g v^{2n-8} (m/M)^2 < 1$. For the marginal case $n=4$,
setting $M=m_{Pl}$ one needs $g < \left(\frac{m_{Pl}}{m}\right)^2 \sim 10^{-70}!$ for solar mass
binary constitutes. We can improve this number by considering
higher dimensional terms, namely larger $n$, but the enhancement
is really minute. Notice that this problem arises at the {\it
classical} level since the coupling to elementary particles is
already too small to represent any trouble. Is only in the
superposition of terms, which build up the massive lump of the
star, that the perturbative expansion breaks down. From this
analysis we conclude that in pure scalar gravity non-derivative self-interactions are extremely
constrained.\\ One could then wonder about more general models including
scalar fields, like tensor-scalar gravity \cite{esposito}. In the
latter in addition to the graviton field a scalar interaction is
added with an action similar to (\ref{ste}) in a curved spacetime
background. Within this type of scenarios the problems we
encountered here can be cured by modifying the power
counting. For instance, a large mass can be added to the scalar
field (larger than the inverse of the solar system distance),
which will render the field a negligible short range interaction. Another
possibility would be to keep it nearly massless but weaken the
coupling to matter to a much feeble strength $M \gg m_{Pl}$. In
this case the 3-scalar diagram (fig 2a) will now scale as $L
\frac{\lambda}{mv^2}\left(\frac{m}{M}\right)^3$. For $\lambda
\sim g M$ one needs $g(m/M)^2 < v^2$ in order to be treat as a
perturbation. By naturalness argument one would expect $g \sim 1$,  and we will then have
a very tiny coupling to elementary particles. For instance, the coupling to a proton will be of the order of
$m_{proton}/M \sim 10^{-60}!$ For a $\phi^4$ theory the 4-scalar diagram would
now scale as $\frac{\tilde\lambda}{v}(m/M)^4$. Compared to the leading
Newtonian potential we get a suppression of order $\tilde\lambda
(m/M)^3(Mrv^2)^{-1}$, which can be seen to be effectively small
for $\tilde\lambda \sim 1$, $M \sim m$. Again the coupling to elementary
particles is highly suppressed. Both solutions will keep the
theory consistent with experimental data, both rely however in the
introduction of a high mass scale into the theory, much higher
than the Planck scale or the scale of particle physics. Perhaps
this is an indication that non-derivative self-interactions are
not present in nature \footnote{The possibility that the dilaton,
the scalar Goldstone mode of the SSB of conformal invariance, could provide an analog of
Einstein theory was recently raised in \cite{raman}. Contrary to the traditional lore
shift invariance does not forbid a non-derivative interaction as in the case of internal symmetries.
A similar fine tunning, which was argued in \cite{raman} to be of the same nature as the cosmological constant,
was invoked to avoid unstable configurations.}

\acknowledgments We would like to thank Walter Goldberger and Ira
Rothstein for helpful comments and discussions. Special thanks to
Christophe Grojean and Francis Bernardeau for organizing such a
great summer school and give us the chance to contribute to this
volume. The work of R.A.P was supported by DOE contracts
DOE-ER-40682-143 and DEAC02-6CH03000.

\end{document}